\documentclass[runningheads]{llncs}
\usepackage{graphicx}
\usepackage{hyperref}
\usepackage{amsmath}
\usepackage{bbold}
\usepackage{mathtools}
\usepackage{booktabs}
\usepackage{multirow}
\usepackage{tabularx}
\usepackage{algorithm}
\usepackage{algorithmic}

\newcounter{protocol}
\newcounter{protocol2}
\newenvironment{protocol}[1]
  {\par\addvspace{\topsep}
   \noindent
   \tabularx{\linewidth}{@{} X @{}}
    \hline
    \stepcounter{protocol}
    \textbf{Protocol \theprotocol} #1 \\
    \hline}
  { \\
    \hline
   \endtabularx
   \par\addvspace{\topsep}}

\begin{document}
\title{Preventing Output-Manipulation in Local Differential Privacy using Verifiable Randomization Mechanism}
\titlerunning{SecureLDP}
%
\author{Fumiyuki Kato\inst{1} \and
Yang Cao\inst{2} \and
Masatoshi Yoshikawa\inst{3} } 
%
%
\institute{Graduate School of Informatics, Kyoto University Dept. of Social Informatics, Japan\\
\email{fumiyuki@db.soc.i.kyoto-u.ac.jp} \and
\email{yang@i.kyoto-u.ac.jp} \and
\email{yoshikawa.masatoshi.5w@kyoto-u.ac.jp}}
\maketitle              
\begin{abstract}
Several randomization mechanisms for local differential privacy (LDP) (e.g., randomized response) are well-studied to improve the utility. 
However, recent studies show that LDP is generally vulnerable to malicious data providers in nature.
Because a data collector has to estimate background data distribution only from already randomized data, malicious data providers can manipulate their output before sending, i.e., randomization would provide them plausible deniability.
Attackers can skew the estimations effectively since they are calculated by normalizing with randomization probability defined in the LDP protocol, and can even control the estimations.
In this paper, we show how we prevent malicious attackers from compromising LDP protocol. 
Our approach is to utilize a verifiable randomization mechanism.
The data collector can verify the completeness of executing an agreed randomization mechanism for every data provider.
Our proposed method completely protects the LDP protocol from output-manipulations, and significantly mitigates the expected damage from attacks.
We do not assume any specific attacks, and it works effectively against general output-manipulation, and thus is more powerful than previously proposed countermeasures.
We describe the secure version of three state-of-the-art LDP protocols and empirically show they cause acceptable overheads according to several parameters.

\end{abstract}
\section{Introduction}
\label{sec:1}

Today’s data science has been very successful in collecting and utilizing large amounts of data. Useful data often includes personal information, and there are serious privacy concerns. 
In particular, recent data breaches \cite{cambridge} \cite{magazine} and strict rules \cite{gdpr} \cite{brazil} by the government greatly promote the concerns.
Local differential privacy (LDP) \cite{DP06} \cite{LDP03} is a promising privacy enhanced technique for collecting sensitive information.
Each client perturbs sensitive data locally by a randomized mechanism satisfying differential privacy, and a server can run analysis such as frequency estimation based on the perturbed data without accessing the raw data. 
We can see the effectiveness and feasibility of LDP in recent production releases of the platformers such as Google \cite{Google14}, Apple \cite{Apple17}, and Microsoft \cite{Microsoft17}, which all utilize LDP for privacy-preserved data curation.

While many studies have been focusing on improving LDP's utility \cite{krr14} \cite{oue17} \cite{ldpicml16} \cite{ldpsigmod19} \cite{ldpsp18} \cite{ldpndss20} in the literature, recent studies \cite{dpat2019} \cite{mail2021} report a vulnerability of LDP protocol and alert the lack of security.
Specifically, \cite{dpat2019} \cite{mail2021} show that the malicious clients can manipulate the analysis such as frequency estimation by sending false data to the collector.
Attackers can skew the estimations effectively by taking advantage of the fact that estimations are calculated by normalizing with randomization probability defined in the LDP protocol, and can even control the estimations.
Their studies significantly highlight the necessity of a secure LDP protocol to defend against malicious clients.
The difficult point of protecting such an attack is that, in a general LDP protocol, others can't verify the results' integrity without the original data because the randomization would provide data providers plausible deniability for their outputs.

Although Cao et al. \cite{dpat2019} showed the some of the countermeasures against malicious clients, their empirical results showed it was not sufficient.
Among their proposed methods, the method of normalizing the estimated probability distribution was shown to be to some extent effective for input-manipulation (i.e., RIA in \cite{dpat2019}).
However, it was not sufficient for output-manipulation (i.e., MGA in \cite{dpat2019}).
In addition, their method for detecting attackers and targeted items depends on the attack method and can only be effective in certain scenarios and cannot be applicable to a wide range of attacks.
They concluded the need for stronger defenses against the attacks.
Concurrently, Cheu et al. \cite{mail2021} also emphasize on the same conclusion for manipulation attacks they call.
There is another promising direction against an attacker who exploits the random mechanism of Differential privacy.
Narayan et al. \cite{vdp2015} propose an interesting scheme to prove integrity for executing correct randomization mechanisms for Differential privacy.
However, since their model is for central DP, it assumes that the data curator, who has the secret data, and the analyst, who creates the proof, do not collude.
This assumption does not hold for LDP, therefore, their proposed method cannot verify valid randomization.

To solve these problems, we propose a novel verifiable LDP protocol based on Multi-Party Computation (MPC) techniques in this work.
Firstly, we specify the attacks of malicious clients have two classes, \textit{output-manipulation} and \textit{input-manipulation} (described in Section \ref{sec:3}).
For input-manipulation attacks, efficient countermeasures have been provided in \cite{dpat2019}, but existing studies cannot prevent output-manipulation sufficiently.
We show the effectiveness of output-manipulation compared to input-manipulation, and highlight the importance of the output-manipulation protection.
To give the protection, we extend Cryptographic Randomized Response Technique (CRRT) \cite{PKC04} to secure the state-of-the-art randomization mechanisms, kRR \cite{krr14}, OUE \cite{oue17} and OLH \cite{oue17}, against output-manipulations.
Our proposed secure protocols do not assume any specific attack, and work effectively against general output-manipulation, and thus is more powerful than previously proposed countermeasures.
The method is based on relatively lightweight cryptographic techniques, hence, the computational overload is acceptable.
To show the fact, we empirically evaluate the performance of our proposed methods.


\noindent
{\bfseries Outline.} 
In Section \ref{sec:2}, we show a background of our work.
In Section \ref{sec:3}, we describe the problem statement.
In Section \ref{sec:4}, we show the details of our proposed method.
In Section \ref{sec:5}, we show the evaluation of our method. 
Finally, we provide the conclusions in Section \ref{sec:6}.

\section{Background: Attacks on LDP protocols}
\label{sec:2}

\subsection{Local Differential Privacy}
\label{sec:2-1}


Differential privacy (DP) \cite{DP06} is a rigorous mathematical privacy definition, which quantitatively evaluates the degree of privacy protection when we publish outputs about sensitive data in a database.
DP is a central model where a trusted server collects sensitive data and releases differentially private statistical information to an untrusted third party.
On the other hand, Local DP (LDP) is a local model, considering an untrusted server that collects clients' sensitive data.
Clients perturb their data on their local environment and send only randomized data to the server for protecting privacy.

In this work, we suppose there is a server $S$ who collects data and aggregates them, and $N$ clients $c_i$ $(0 \le i \le N-1)$ who send their sensitive data in a local-differentially private manner.
Each client has a item $v$ which is a categorical data, and the items have $d$ domains and $v \in [0,d-1](\coloneqq[d])$, additionally, $v_i$ denotes $c_i$'s item.
The clients randomize their data by randomization mechanism $\mathcal{A}$, and $c_i$ send $\mathcal{A}(v_i) = y_i (\in D)$ to the server, where $D$ is output space of $\mathcal{A}$.
The server estimates some statistics by $\mathcal{F}(y_0, ..., y_{N-1})$.
In particular of this work, $\mathcal{F}_k$ corresponds to \textit{frequency estimation} for item $k$ (i.e., how many clients have chosen item $k$).
The formal LDP definition is as follows:
\begin{definition}[$\epsilon$-local differential privacy ($\epsilon$-LDP)]
\label{def:ldp}
A randomization mechanism $\mathcal{A}$ satisfies $\epsilon$-LDP, if and only if for any pair of input values $v, v' \in [d]$ and for all randomized output $y \in D$, it holds that
\begin{equation}
\nonumber
  \Pr[\mathcal{A}(v)\in y] \leq e^{\epsilon} \Pr[\mathcal{A}(v')\in y].
\end{equation}
\end{definition}

Under a specific randomized algorithm $\mathcal{A}$, we want to estimate the frequency of any items.
Wang et al. \cite{oue17} introduce "pure" LDP protocols with nice symmetric property and a generic aggregation procedure to calculate unbiased frequency estimations from given randomization probabilities.
Let \textsf{Support} be a function that maps each possible output $y$ to a set of input that $y$ supports.
\textsf{Support} is defined for each LDP protocol and it specifies how the estimation can be computed under the LDP protocol.
A formal definition of pure LDP is as follows:
\begin{definition}[Pure LDP \cite{oue17}]
\label{def:pure}
A protocol given by $\mathcal{A}$ and \textsf{Support} is pure if and only if there exist two probability values $p > q$ such that for all $v_1$,
\begin{align}
\nonumber
  &\Pr[\mathcal{A}(v_1)\in \{o | v_1 \in \mathsf{Support}(o) \}] = p, \\
  \forall_{v_2\neq v_1}&\Pr[\mathcal{A}(v_2)\in \{o | v_1 \in \mathsf{Support}(o) \}] = q.
\end{align}
where $p, q$ are probabilities, and $q$ must be the same for all pairs of $v_1$ and $v_2$.
\end{definition}
While maximizing $p$ and minimizing $q$ make the LDP protocol more accurate, under $\epsilon$-LDP it must be $\frac{p}{q} \le e^{\epsilon}$.
The important thing is that, in pure LDP protocol, we can simply estimate the frequency of item $k$ as follows:
\begin{equation}
\label{eq:freq}
\mathcal{F}_k = \cfrac{\sum_{i}\mathbb{1}_{\mathsf{Support}(y^i)}(k)-Nq}{p-q}
\end{equation}
We can interpret that this formula normalizes observed frequencies using probabilities $p$ and $q$ to adjust for randomization.

For frequency estimation under LDP, we introduce three state-of-the-art randomization mechanisms, kRR \cite{krr14}, OUE \cite{oue17} and OLH \cite{oue17}.
These mechanisms includes three steps: (1) \textsf{Encode} is encoding function: $\mathcal{E}: v (\in [d]) \rightarrow v'(\in [g])$ , (2) \textsf{Perturbation} is randomized function: $\mathcal{A}: v' (\in [g]) \rightarrow y (\in D)$ , (3) \textsf{Aggregation} calculates estimations from all collected values: $\mathcal{F}: (y_{0}, ... ,y_{N-1}) \rightarrow \mathbb{R}$.
Formal proofs that each protocol satisfies $\epsilon$-LDP can be found in \cite{oue17}.

\vspace{4px}
\noindent
\textbf{k-ary Randomized Response (kRR)} is an extension of Randomized Response \cite{randomizedresponse} to meet $\epsilon$-LDP.
In particular, kRR provides accurate results in small item domains.
This mechanism does not require any special encoding, and provides a identity mapping $\mathcal{E}(v) = v$ $([g]=[d])$.
Perturbation is as follows;
\begin{equation}
Pr[\mathcal{A}(v) = y] = 
\begin{cases}
    p=\cfrac{e^{\epsilon}}{e^{\epsilon}+d-1} \;, &\mathrm{if} \; y = v \\
    q=\cfrac{1-p}{d-1}=\cfrac{1}{e^{\epsilon}+d-1} \;,        &\mathrm{if} \; y\neq v
\end{cases}
\end{equation}

\noindent
For aggregation, we can consider \textsf{Support} function as $\mathsf{Support}(v) = (v)$ and make this follow pure LDP protocol (Def. \ref{def:pure}).
Therefore, aggregation follows Eq.(\ref{eq:freq}).

\vspace{4px}
\noindent
\textbf{Optimized Unary Encoding (OUE)} encodes item $v$ into $d$-length bit vector and encode function is defined as $\mathcal{E}(v) = [0,...,0,1,0,...,0]$ where only single bit corresponding to $v$-th position is 1.
Final output space is also $d$ dimensional bit vector $\{0,1\}^{d}$ (e.g. $\mathbf{y}=[1,0,1,1,0]$).
Let $i$-th bit of output vector as $y_i$, perturbation is as follows; 
\begin{equation}
\label{eq:oue}
Pr[y_i = 1] = 
\begin{cases}
    p=\cfrac{1}{2}\; , &\mathrm{if} \; i = v \\
    q=\cfrac{1}{e^{\epsilon}+1}\; , &\mathrm{if} \; i\neq v
\end{cases}
\end{equation}
These $p$ and $q$ minimize the variance of the estimated frequency in similar bit vector encoding (e.g. RAPPOR \cite{Google14}).
In aggregation step, we consider \textsf{Support} function as $Support(\mathbf{y})=\{v|y_v=1\}$, and also calculate using Eq.(\ref{eq:freq}).

\vspace{4px}
\noindent
\textbf{Optimized Local Hashing (OLH)} employs hash function for dimensional reduction to reduce communication costs.
It picks up $H$ from a universal hash function family $\mathbb{H}$, and $H$ maps $v \in [d]$ to $v' \in [g]$ where $2 \le g < d$.
Therefore, encode function is $\mathcal{E}(v) = H(v)$.
Perturbation is the same as kRR, except that the input/output space is $[g]$.
Then, $p$ and $q$ is defined as follows;
\begin{equation}
Pr[\mathcal{A}(x) = y] = 
\begin{cases}
    p=\cfrac{e^{\epsilon}}{e^{\epsilon}+g-1} \;, &\mathrm{if} \; y = H(v) \\
    q=\cfrac{1}{g}\cdot p + \left(1-\cfrac{1}{g}\right) \cdot \cfrac{1}{e^{\epsilon}+g-1} =\cfrac{1}{g} \;, &\mathrm{if} \; y \neq H(v)
\end{cases}
\end{equation}

\noindent
In aggregation step, we consider \textsf{Support} function as $Support(\mathbf{y})=\{v|v\in [d] \mbox{ and } y = H(v)\}$ and follow Eq.(\ref{eq:freq}) using $p$ and $q$.

\subsection{Attacks on LDP protocols}
\label{sec:2-2}
In this subsection, we introduce two important studies suggesting caution to necessity of secure LDP protocols.

\vspace{2px}
\noindent
\textbf{Targeted Attack.}
Cao et al. \cite{dpat2019} focus on \textit{targeted} attacks to LDP protocols, where the attacker tries to promote the estimated frequencies of a specific item set.
Considering the attacker against the LDP protocols, $M$ malicious clients, who can arbitrarily control local environments and send crafted data to the server, are injected by the attacker. (They call \textit{data poisoning attacks}.)
Attacker wants to promote $r$ target items $T = \{t_1,..., t_r\}$ in the frequency estimation.
Cao et al. propose three attacks: Random perturbed-value attack (RPA), Random item attack (RIA), Maximal gain attack (MGA).
The first two attacks are as baselines and MGA is an optimized attack.
In RIA, malicious clients perform uniform random samplings of a value from the target item set. 
And then, following the LDP protocol, encoding and perturbation are performed and sent to the server.
MGA is more complicated than others.
It aims to maximize the attacker's overall gain $G$: sum of the expected frequency gains for the target items, $G=\sum_{t\in T}\mathbb{E}[\Delta f_t]$ where $\Delta f_{t}$ represents the increase of estimated frequency of item $t$ $(\forall t \in T)$ from without attack to with attack.
In MGA, the output item selection is performed according to the optimal solution maximising the attacker's gain, and sent to the server without perturbation. 

Cao et al. describe the details of these three attacks against kRR, OUE, OLH in the frequency estimation and give theoretical analysis.
The summary of the results is shown in Table \ref{tbl:targeted}.
The table shows the overall gains of the three attacks against kRR, OUE, and OLH.
MGA can achieve the highest gains for all protocols, clearly because MGA maximizes the gains.
A notable point is a difference, summarized in Table \ref{tbl:summary_two} that shows the difference of gains between MGA and RIA.
They respectively correspond to \textit{output-manipulation} and \textit{input-manipulation} (described later) in our work.
Note that the difference is remarkable, especially under the higher privacy budget.


\vspace{2px}
\noindent
\textbf{Untargeted Attack.}
Albert et al. \cite{mail2021} analyze manipulation attacks in LDP.
Compared to Cao et al.'s work, their study mainly focuses on \textit{untargeted} attacks, where the attackers aim to skew the original distribution and degrade the overall estimation accuracy of the server.

They suggest for the LDP protocols that the architecture is inherently vulnerable to malicious clients' manipulations.
They suppose a general manipulation attack: the attacker injects $M$ users in $N$ clients in the LDP protocol and these injected users can send arbitrary data sampled from carefully skewed distributions to the server without supposed perturbation.
We consider this attacker model corresponds to MGA in \cite{dpat2019} and \textit{output-manipulation} (described later) in this paper.
One of their contribution we should focus on is that they show the general manipulation attack can skew the estimated distribution by $\Omega(\frac{M\sqrt{d}}{\epsilon N})$ in the frequency estimation, which cause larger error than input-manipulation by about a $\frac{\sqrt{d}}{\epsilon}$ factor (Table \ref{tbl:summary_two}).
The difference is, for example, defined as $l_1$-norm of the original and skewed distribution.

\begin{table*}[t]
    \centering
    \begin{tabular}{lccc}
    \toprule
     & kRR & OUE & OLH  \\
    \midrule
    RPA (output-manipulation)
    & $\beta(\frac{r}{d} - f_T)$
    & $\beta(r - f_T)$
    & $-\beta f_T$ \\
    RIA (input-manipulation)
    & $\beta(1 - f_T)$
    & $\beta(1 - f_T)$
    & $\beta(1 - f_T)$ \\
    MGA (output-manipulation)
    & $\beta(1 - f_T) + \frac{\beta(d-r)}{e^{\epsilon}-1}$
    & $\beta(2r - f_T) + \frac{2\beta r}{e^{\epsilon}-1}$
    & $\beta(2r - f_T) + \frac{2\beta r}{e^{\epsilon}-1}$ \\
    \bottomrule
    \end{tabular}
\caption{MGA can achieve the highest \textbf{gains} against all three protocols. $\beta=\frac{M}{N+M}$ and $f_T=\sum_{t\in T} f_t$ in the table. (The summary results of \cite{oue17}.)}
\label{tbl:targeted}
\vspace{-3mm}
\end{table*}
\begin{table*}[t]
    \centering
    \addtolength{\leftskip} {-2cm}
    \addtolength{\rightskip}{-2cm}
    
    \begin{tabular}{lccc}
    \toprule
    \multirow{2}{*}{\textbf{Targeted Attack \cite{dpat2019}}} & kRR   & OUE   & OLH   \\ \cline{2-4} 
                                     & $+\left(\cfrac{\beta(d-r)}{e^{\epsilon}-1}\right)$      & $+\left(\beta(2r - 1) + \cfrac{2\beta r}{e^{\epsilon}-1}\right)$      & $+\left(\beta(2r - 1) + \cfrac{2\beta r}{e^{\epsilon}-1}\right)$       \\ \midrule
    \textbf{Untargeted Attack \cite{mail2021}}               & \multicolumn{3}{c}{$\times \Omega\left(\cfrac{\sqrt{d}}{\epsilon}\right)$} \\ \bottomrule
    \end{tabular}
\vspace{5px}
\caption{Overall, output-manipulations are much more vulnerable than input-manipulation. The differences of both manipulations gain are calculated by output-manipulation gain $-$ input-manipulation gain (resp. output-manipulation gain $/$ input-manipulation gain) in Targeted (resp. Untargeted) Attack.}
\label{tbl:summary_two}
\vspace{-7mm}
\end{table*}

\noindent
\textbf{Summary.}
We summarize these notable results in Table \ref{tbl:summary_two}, showing that how effective output-manipulation can attack compared to input-manipulation.
The above two previous studies' common conclusion is highlighting the great necessity of enforcing the correctness of users' randomization to defend the output-manipulation attacks.

\section{Problem statements}
\label{sec:3}

Firstly, we give some notations to LDP protocols, partially following the abovementioned in Section \ref{sec:2}.
We denote a single LDP protocol as $\pi_i$, where a client $c_i$ sends a sensitive data $v$ to server $S$ in $\epsilon$-LDP manner.
Encode and perturbation are denoted together as $\phi$.
$\phi$ is a probabilistic function (i.e., randomization mechanism) that takes $v\in [d]$ as input and output $y\in D$, such that output space $D=[d]$ if kRR, $D=\{0,1\}^{d}$ if OUE, $D=[g]$ if OLH.
And we denote overall protocol including all clients as $\Pi=\{\pi_i| i\in [N]\}$.

\subsection{Overview of our goal}

\begin{figure}[t]
    \centering
    \includegraphics[width=0.95\hsize]{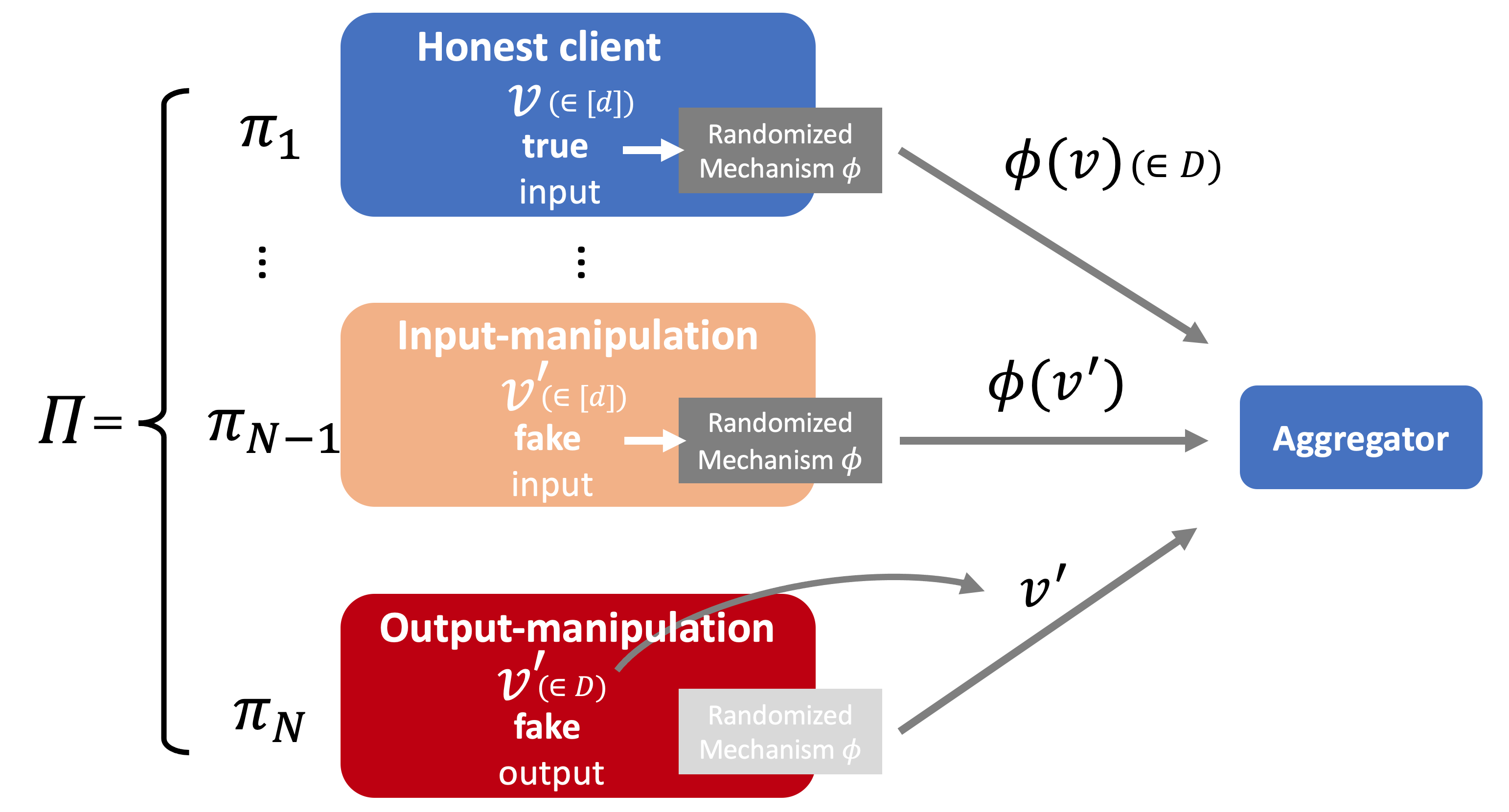}
    \caption{From top to bottom, normal protocol, input-manipulation attack and output-manipulation attack against an LDP protocol.}
    \label{fig:two_attacks}
    \vspace{-3mm}
\end{figure}

An attacker against $\Pi$ injects compromised users into the protocol to send many fake data to a central server.
Note that such an attack results in manipulation against a single protocol $\pi$ by each compromised user.
Therefore, we consider security for $\pi$, and by protecting security for $\pi$, we can naturally protect security for $\Pi$.
As for the attacker's capability, the attacker can access the implementation of $\phi$ because this is executed on clients' local, and he knows all parameters and functions including $\phi$, $\epsilon$, $d$, $D$ and $Support(y)$, and employs this information to craft effective malicious outputs.
However, in fact, there is little variation in the attacker's behavior because the server can easily deny the protocol if output $y\notin D$.
Under such conditions, as shown in Figure \ref{fig:two_attacks}, we can observe that an attacker can carry out the following two classes of attacks:

\vspace{2px}
\noindent
\textbf{Input-manipulation} supposes that the attacker can only select input data $v\in [d]$ and cannot interfere with other parameters and functions in $\pi$ (middle in Figure \ref{fig:two_attacks}).
In other words, the attacker must send $y=\phi(v)$.
But in a realistic setting, we should consider it a too strong assumption, as it allows an attacker to have complete control over the local system environment.
For example, in targeted attack, RIA corresponds to this class of attacks.

\vspace{2px}
\noindent
\textbf{Output-manipulation} supposes the attacker can send arbitrary outputs to the server (bottom in Figure \ref{fig:two_attacks}).
This corresponds the attacker can ignore all parameters and functions $\epsilon$, $\phi$ by manipulating outputs directly.
This attack is an entirely reasonable attack against a general LDP protocol because the server cannot distinguish between true data or fake data.
In targeted attack, it corresponds to MGA or the attacks proposed in \cite{mail2021} for untargeted attack.
Generally, this class effectively attacks, as shown in Table \ref{tbl:summary_two}.

An important observation from Section \ref{sec:2-2} is that input-manipulation is much less effective than output-manipulation.
Therefore, the natural direction is to defend against output-manipulation and limit the attack to the range of input-manipulation to achieve secure LDP protocols.
On the other hand, it is hard to prevent input-manipulation completely.
These have been studied in the fields of game theory \cite{basian04} \cite{game17} or truth discovery \cite{truthdiscovery14}, and we leave such a solution as future work.


Overall, our goal is to mitigate attacks against the LDP protocols by completely defending output-manipulation and limiting to input-manipulation.
For this purpose, we consider enforcing the correct mechanism $\phi$ for protocol $\pi$.
The key idea is to make the protocol verifiable against malicious clients from a server.
In the rest of the paper, we refer to this property as $\textit{output-manipulation secure}$. (It is also expressed simply as $secure$, and we call it as secure LDP protocol.)
\begin{definition}[output-manipulation secure]
\label{def:omsecure}
An LDP protocol $\pi$ is output-manipulation secure if any malicious client cannot perform output-manipulation and can only perform input-manipulation against $\pi$.
\end{definition}

\subsection{Security definitions}
In this subsection, we clarify what we should achieve for a secure LDP protocol.
Similar to \cite{PKC04}, security definitions of secure LDP protocol are consistent with a traditional secure two-party computation (2PC) protocol described in \cite{twopc}.
It considers an ideal world where we can employ Trusted Third Party (TTP) to execute arbitrary confidential computations surely.
And we aim to replace the TTP with a real-world implementation of cryptographic protocol $\pi=(c, S)$ between client $c$ and server $S$.
The protocol's flow when using a TTP is very simple.
The client $c$ sends input $v$ to the TTP, and the TTP provides $y=\phi(v)$ to $S$.
After all, $c$ and $S$ never receive any other information; $S$ does not know $v$, and $c$ does not know $y$.
($S$ can estimate $v$ from $y$ and $\phi$, but $c$'s privacy should be guaranteed by LDP.)

While it is seemingly obvious that this ideal world's protocols will satisfy our requirements, let us review possible attacks closely.
Goldreich \cite{twopc} summarizes that there are just three types of attacks in malicious model 2PC against ideal world protocols;
(1) denial of participation in the protocol;
(2) fake input, not the true one;
(3) aborting the protocol prematurely.
We cannot hope to avoid these, but (1) and (3) cannot influence of the estimation of original data distributions in LDP protocols.
(2) is exactly input-manipulation described in the previous subsection.
Thus, it is sufficient that the ideal world in 2PC is \textit{output-manipulation secure} (see Def. \ref{def:omsecure}) in the LDP protocols.

Considering the substituted cryptographic protocol $\pi=(c, S)$, let client $c$ as prover $\mathcal{P}$ and server $S$ as verifier $\mathcal{V}$ and $\pi=(\mathcal{P}, \mathcal{V})$.
More specifically, we should guarantee secure LDP under the worst case that both $\mathcal{P}$ and $\mathcal{V}$ behave maliciously.
The case where $\mathcal{P}$ is malicious is obvious, considering output-manipulation.
Still, in the $\mathcal{V}$'s case, it is because, given the original scenario of LDP, we need to guarantee the privacy of $\mathcal{P}$.
And, we assume $\mathcal{P}$ is a polynomial computational adversary and $\mathcal{V}$ is unbounded.

Following \cite{PKC04}, the ideal world protocol can be substituted with protocol $\pi$ if
(a) for any prover algorithm $\mathcal{P}^*$, $\mathcal{V}$ who receives $\phi(v)=y$ accepts only when $\mathcal{P}^*$'s secret input is surely $v$, or otherwise halts with negligible error; (b) for any prover algorithm $\mathcal{P}^*$, $y$ is indistinguishable from other categories; (c) for any verifier algorithm $\mathcal{V}^*$, $v$ is indistinguishable from other categories.
Additionally, we need to verify that the randomization function $\phi$ used in the protocol does indeed satisfy $\epsilon$-LDP.

Let $\mathrm{view}_{\mathcal{P}}$ (resp. $\mathrm{view}_{\mathcal{V}}$) as the set of messages generated by the protocol that $\mathcal{P}$ (resp. $\mathcal{V}$) can observe, and let $k$ as a security parameter that increase logarithmically with cryptographic strength.
Then, our security definitions are reduced as following three properties:
\vspace{-2mm}
\begin{itemize}
    \item \textbf{Verifiability}:
    This property corresponds to above-mentioned (a).
    We consider the protocol is verifiable if it satisfies as follows;
    \begin{equation}
    \label{eq:com}
    Pr[ \mathcal{V} \mbox{ does not halts } | y \leftarrow \phi(*) ] = 1 \mbox{ and, }
    \end{equation}
    \vspace{-6mm}
    \begin{equation}
    \label{eq:sou}
    1 - Pr[ \mathcal{V} \mbox{ halts } | y \leftarrow \mathcal{P^*} ] < negl(k)
    \end{equation}
    where $negl(k)$ is negligible function in $k$, $y \leftarrow \phi(*)$ means $y$ is obtained by correct execution of $\phi$ and $y \leftarrow \mathcal{P^*}$ means $y$ is obtained by $\mathcal{P^*}$ other than correct $\phi(v)$.
    
    \item \textbf{Indistinguishability}: 
    This property corresponds to (b) and (c).
    (b) satisfies if  $\mathrm{view}_{\mathcal{P}^*}$ has indistinguishable distributions for any input category $v \in [d]$.
    Formally, we define this property as follows;
    for any adversary $\mathcal{P^*}$, 
    \begin{equation}
    \label{eq:ind_p}
    |Pr[\mathcal{P^*}(\mathrm{view}_{\mathcal{P}^*}, v) = y] - Pr[\mathcal{P^*}(v) = y]| < negl(k)
    \end{equation}
    where $negl(k)$ is negligible function in $k$. 
    This means that a malicious client can use any information obtained from the protocol but only get negligible information about the final output of the server side.
    Similary, (c) satisfies if, for any unbounded adversary $\mathcal{V^*}$, 
    \begin{equation}
    \label{eq:ind_v}
    |Pr[\mathcal{V^*}(\mathrm{view}_{\mathcal{V}^*}, y) = v] - Pr[v|y]| < negl(k)
    \end{equation}
    
    \item \textbf{Local Differential Privacy}: The randomization mechanism $\phi$ in the given protocol must satisfy $\epsilon$-LDP as shown in Def. \ref{def:ldp}.
    The verification of the correct execution is performed in Eq. (\ref{eq:com}).
\end{itemize}

\section{Proposed Method}
\label{sec:4}
We design secure LDP protocols for kRR, OUE, and OLH, respectively, so that we defend output-manipulations completely.
In our method, a major building block is Cryptographic Randomized Response Technique (CRRT) \cite{PKC04} which employs Pedersen's commitment scheme \cite{ped91} for secure verifiability using additive homomorphic property, and Naor-Pinkas 1-out-of-n Oblivious Transfer (OT) technique \cite{nbo01} for tricks for a verifiable randomization mechanism.
Overall, the proof of validity is based on disjunctive proof \cite{dis94}.
It is a lightweight interactive proof protocol based on a secret sharing scheme and, can perform witness-indistinguishable \cite{witness} proofs of knowledge (similar to zero-knowledge proofs).
Combined with the security of the encryption scheme proposed in \cite{PKC04}, it is possible to securely prove that the output value $y$ is obtained by sampling from a probability distribution that satisfies the $\epsilon$-LDP i.e. $y=\phi(v)$.
For simplicity, we explain several phases separately in the following protocol description (Protocol \ref{pro:krr}, \ref{pro:oue}), but they can be done simultaneously in the actual implementation.

Before explanation of the protocols in detail, we introduce the following cryptographic setting.
Assume that $p$ and $q$ are sufficiently large primes such that $q$ divides $p-1$, $\mathcal{Z}_{p}$ has a unique subgroup $\rm{G}$ of order $q$.
$q$ is the shared security parameter between $\mathcal{P}$ and $\mathcal{V}$.
Security parameter $k$ is $k=\log_{2}{q_{max}}$ such that $q_{max}$ is the maximum value of possible $q$.
We select $g$ and $h$  as a public key.
They are two generators of $\rm{G}$ and, their mutual logarithms $\log_{g}{h}$ and $log_h{g}$ are hard to compute.
We use this public key in the following protocols.

\subsection{Secure kRR}
Protocol \ref{pro:krr} shows the details of the secure version of kRR, which is basically extension of CRRT \cite{PKC04} to satisfy LDP for multidimensional data.
As a whole, in the setup phase, both $\mathcal{P}$ and $\mathcal{V}$ prepare the same parameters $l, n, z$ from accuracy parameter $width$ and privacy budget $\epsilon$ by Algorithm \ref{alg:params_krr}.
$l, n, z$ identify an categorical probability distributions that satisfies LDP and we use it in 1-out-of-$n$ OT for verifiable random sampling.
In mechanism phase, $\mathcal{P}$ creates a vector $\boldsymbol{\mu}$ representing the categorical distribution containing $n$ data where each data $\mu_i$ corresponds to one of the categories $[d]$.
$width$ (i.e., $n$) is the size of the vector and decides a trade-off between accuracy to approximate LDP and overheads caused by the protocol.
For proof P2, we use $z^{\mu_i}$ instead of $\mu_i$.
All $z^{\mu_i}$ is encrypted to $y_i$ by an encryption scheme that combines Pedersen's commitment and OT.
Only the $\mu_{\sigma}$, where $\sigma$ is pre-chosen by $\mathcal{V}$, can be decrypted correctly.
Such a trick allows us to surely perform random sampling from vector $\boldsymbol{\mu}$ representing the categorical distribution.
In the proof phase, two proofs are verified in the protocol.
The first one is a disjunctive proof that for each encrypted data $y_i$ belonging to one of the categories $[d]$ (\textbf{P1}) .
The second one also uses a disjunctive proof that the summation of the vector used as categorical distribution in the OT belongs to one of the possible values (\textbf{P2}).
There are just $d$ possible values for the summation of $\boldsymbol{\mu}$ (4.(a)).

\begin{algorithm}[t]
\caption{\textsc{DecideSharedParameters}}
\label{alg:params_krr}
\begin{algorithmic}[1]
\renewcommand{\algorithmicrequire}{\textbf{Input:}}
\renewcommand{\algorithmicensure}{\textbf{Output:}}
\newcommand{\algorithmicbreak}{\textbf{break}}
\newcommand{\BREAK}{\STATE \algorithmicbreak}
\REQUIRE $\epsilon$, $width$ \\
\STATE $i \leftarrow \lfloor\frac{e^{\epsilon}}{(d-1) + e^{\epsilon}}\rfloor$ // as an integer
\WHILE{$i > 0$}
\IF{$(width - i)$ divides $(d-1)$}
\STATE $g \leftarrow gcd(i, width, \frac{width-i}{d-1})$
\STATE $l, n \leftarrow \frac{i}{g}$, $\frac{width}{g}$
\BREAK
\ENDIF
\STATE $i \leftarrow i-1$
\ENDWHILE
\STATE $z \leftarrow \max ([l, \frac{n-l}{d-1}]) + 1$
\ENSURE $l$, $n$, $z$
\end{algorithmic} 
\end{algorithm}

Here, we confirm Protocol \ref{pro:krr} is secure.
From the protocol, prover and verifier get $\mathrm{view}_{\mathcal{P}}=\{g^a, g^b, g^{ab-\sigma-1}, x_i, x\}$ and $\mathrm{view}_{\mathcal{V}}=\{w_i, y_i, com_i^{(j)}, c_i^{(j)}, h_i^{(j)}, com_j, c_i, h_i\}$ for all $i\in [n], j\in [d]$ respectively.

Firstly, we consider indistinguishability.
Our encryption scheme (e.g. $\mu_i$ is encrypted to $y_i$) is the same as the one presented in \cite{PKC04}, which have been shown that a protocol using the scheme is sufficiently indistinguishable for $\mathcal{P}^*$ and $\mathcal{V}^*$.
That is, it is as hard for $\mathcal{P}^*$ to know about the $\sigma$, and also hard for $\mathcal{V}^*$ to guess the distribution of $\boldsymbol{\mu}$ and input $v$.
Considering the attacker views, for $\mathcal{P}^*$, calculating $\sigma$ from $\mathrm{view}_{\mathcal{P}}$ is as hard as the Decisional Diffie Hellman (DDH) problem.
And $x$ and $x_i$ are completely random integers.
For $\mathcal{V}^*$, $(w_i, y_i)$ of $\mathrm{view}_{\mathcal{V}}$ is indistinguishable by the security of the cryptographic scheme, and $(com_i^{(j)}, c_i^{(j)})$ is also indistinguishable because of the secret sharing scheme \cite{dis94}.

Verifiability is satisfied by proofs, P1 and P2.
If both P1 and P2 are verified, $\mathcal{V}$ itself selects one value from the verified vector by OT.
Then, for any operation by $\mathcal{P^*}$, $\mathcal{V}$ can confirm the correctness of the protocol.
Hence, verifiability entirely depends on the protocol that proves the P1 and P2.
We use disjunctive proofs and Eq.(\ref{eq:com}) and Eq.(\ref{eq:sou}) are respectively satisfied by the completeness and soundness of the disjunctive proofs shown in \cite{dis94}.

Lastly, Algorithm \ref{alg:params_krr} definitely generates $l, n$ such that $\frac{l}{n} \le \frac{e^{\epsilon}}{(d-1) + e^{\epsilon}}$ and $\frac{n-l}{d-1} \ge \frac{1}{(d-1) + e^{\epsilon}}$.
Hence, because random sampling from $\boldsymbol{\mu}$ is equivalent to kRR with $p=\frac{l}{n}, q=\frac{n-l}{d-1}$, at least $\epsilon$-LDP is satisfied.

\subsection{Secure OUE}
We show the secure version of OUE protocol in Protocol \ref{pro:oue}.
Unlike kRR, OUE sends a $d$-length bit vector where each $i$-th bit corresponds that client likely has the item $i\in [d]$.
In OUE, mechanism $\phi$ performs random bit flips with given constant probability independently for each bit.
The Bernoulli distributions, which determines the probabilities of each flip, are approximated by a distribution of $n$-length bit vectors, and as in the case of kRR, verifiable random sampling is achieved by a trick using Pedersen's commitment and OT.
However, there are $d$ distribution vectors since it needs for each category.
In addition, each vector's distribution is one of two types: $j$-th vector such that secret input $v=j$ or otherwise ( i.e., $p$ or $q$ in Eq. (\ref{eq:oue})).
Thus, we perform independent OT and decide 0 or 1 for $d$ categories and finally, get randomized output $[\mu_{\sigma_1},...,\mu_{\sigma_d}]$.

\begin{protocol}{secure kRR}
\refstepcounter{protocol2}
\label{pro:krr}
Client $c$ as prover $\mathcal{P}$ who holds an secret input $v\in [d]$ and server $S$ as verifier $\mathcal{V}$. $\epsilon$ is privacy budget and $width$ is a parameter representing the degree of approximation.
\begin{enumerate}
  \item \textbf{Setup phase.}
  \begin{enumerate}
    \item
    $\mathcal{P}$ and $\mathcal{V}$ run \textsc{DecideSharedParameters}$(\epsilon, width)$ and prepare $l, n, z$ as shown in Algorithm \ref{alg:params_krr}.
    This is an algorithm for approximating integers $l, n, z$ for given $\epsilon$ with as little degradation in accuracy as possible while still satisfying privacy protection.
    
    \item
    $\mathcal{V}$ selects $\sigma \in [n]$.
    And $\mathcal{P}$ prepares a $n$-length random number vector $\boldsymbol{\mu}=(\mu_1, ... ,\mu_n)$ where for all $1\le i \le n$, $\mu_i \in [d]$, the vector satisfies $\#\{\mu_i | \mu_i \in \boldsymbol{\mu} \mbox{ and } \mu_i = v\} = l$ and for all $\{v'| v'\in [d] \setminus \{v\} \}$, $\#\{\mu_i | \mu_i = v' \} = \frac{n-l}{d-1}$ where $\#\{\cdot\}$ returns count on a set.
    
  \end{enumerate}
  
  \item \textbf{Mechanism phase.}
  \begin{enumerate}
    \item
    $\mathcal{V}$ picks random $a, b \leftarrow \mathbb{Z}_q$ and sends $g^a$, $g^b$ and $g^{ab-\sigma+1}$ to $\mathcal{P}$.
    
    \item
    For all $i\in \{1,...,n\}$, $\mathcal{P}$ performs the following subroutine;
    (1) Generate $(r_i, s_i)$ at random;
    (2) Compute $w_i \leftarrow g^{r_i}(g^{a})^{s_i} = g^{r_i + as_i}$ and $h_i \leftarrow (g^b)^{r_i}(g^{ab-\sigma+1}g^{i-1})^{s_i} = g^{(r_i+as_i)b+(i-\sigma)s_i}$;
    (3) Encrypt $\mu_i$ to $y_i$ as $y_i \leftarrow g^{z^{\mu_i}}h^{_i}$.
    Then, send ($w_i$, $y_i$) to $\mathcal{V}$.
    
    \item
    $\mathcal{V}$ computes $w_{\sigma}^b$ where $\sigma$ is what $\mathcal{V}$ choose at setup phase, and computes $g^{\mu_{\sigma}} \leftarrow \frac{y_{\sigma}}{h^{w_{\sigma}^b}}$.
    And then, find $\mu_{\sigma}$ from the result and $g$.
    Thus, $\mathcal{V}$ receives $\mu_{\sigma}$ as a randomized output from $\mathcal{P}$.
    
  \end{enumerate}
  \item \textbf{Proof phase for P1.}
  \begin{enumerate}
    \item 
    For all $j \in [d] \setminus \{\mu_i\}$, for all $i \in \{1,...,n\}$, $\mathcal{P}$ generates challenge $c_i^{(j)}$ and response $s_i^{(j)}$ from $\mathbb{Z}_q$ and prepares commitments $com_i^{(j)} \leftarrow h^{s_i^{(j)}}/(y_i / g^{z^j})^{c_i^{(j)}}$.
    For $\{\mu_i\}$ and for all $i \in \{1,...,n\}$, $\mathcal{P}$ generates $w_i \leftarrow \mathbb{Z}_q$ and let $com_i^{(\mu_i)} = h^{w_i}$.
    Then, send $com_i^{(j)}$ to $\mathcal{V}$, for all $i, j$.
    
    \item
    $\mathcal{V}$ picks $x_i \leftarrow \mathbb{Z}_q$ for all $i \in \{1,...,n\}$ and sends it to $\mathcal{P}$.
    
    \item
    For all $i \in \{1,...,n\}$, $\mathcal{P}$ computes $c_i^{(\mu_i)} = x_i - \sum_{j \in [d] \setminus \mu_i}{c_i^{(j)}}$ and $s_i^{(\mu_i)} = v_ic_i^{(\mu_i)} + w_i$.
    Then, send $c_i^{(j)}$ and $s_i^{(j)}$ for all $i, j$ to $\mathcal{V}$.
    
    \item
    Finally, $\mathcal{V}$ checks if $h^{s_i^{(j)}} = b(y_i/g^{z^j})^{c_i^{(j)}}$ for all $j\in [d]$ and $x_i = \sum_{j\in [d]}{c_i^{(j)}}$, for all $i\in {1,...,n}$. Otherwise halts.
    
  \end{enumerate}
  
  \item \textbf{Proof phase for P2.}
  \begin{enumerate}
    \item 
    For all $j \in [d] \setminus \{v\}$, $\mathcal{P}$ generates challenge $c_j$ and response $s_j$ from $\mathbb{Z}_q$ and prepares commitments $com_j \leftarrow h^{s_j}/(\prod_{i\in \{1,..,n\} y_i / g^{Z_j})^{c_j}}$ where $Z_j = \frac{n-l}{d-1}\left(\sum_{k \in [d] \setminus \{j\}} z^k \right) + lz^j$.
    And $\mathcal{P}$ generates $w \leftarrow \mathbb{Z}_q$ and let $com_v = h^{w}$.
    Then, send $com_j$ to $\mathcal{V}$, for all $j\in [d]$.
    
    \item
    $\mathcal{V}$ picks $x \leftarrow \mathbb{Z}_q$ and sends it to $\mathcal{P}$.
    
    \item
    $\mathcal{P}$ computes $c_v = x - \sum_{j \in [d] \setminus \{v\}}{c_j}$ and $s_v = \left(\sum_{i \in {1,...,n}}v_i\right)c_v + w$.
    Then, send $c_j$ and $s_j$ for all $j$ to $\mathcal{V}$.
    
    \item
    Finally, $\mathcal{V}$ checks if $h^{s_j} = b(\prod_{i\in \{1,..,n\}} y_i / g^{Z_j})^{c_j}$ for all $j\in [d]$ and $x = \sum_{j\in [d]}{c_j}$. Otherwise halts.
  \end{enumerate}
\end{enumerate}
\end{protocol}

\begin{protocol}{secure OUE}
\refstepcounter{protocol2}
\label{pro:oue}
$\mathcal{P}$, $v\in [d]$, $\mathcal{V}$, $width$, $\epsilon$ as with Protocol \ref{pro:krr}.
\begin{enumerate}
  \item \textbf{Setup phase.}
  \begin{enumerate}
    \item
    $\mathcal{P}$ and $\mathcal{V}$ set $l, n$ as $\lceil\frac{1}{1 + e^{\epsilon}}\cdot width\rceil$ and $width$ itself respectively.
    
    \item
    $\mathcal{V}$ selects $d$ random numbers ${\boldsymbol{\sigma}}=\{\sigma_1,...,\sigma_d\}$ where $1\le \sigma_j \le n$.
    $\mathcal{P}$ prepares $d$ $n$-length random bit vectors $\vec{\boldsymbol{\mu}}=(\boldsymbol{\mu_1},...,\boldsymbol{\mu_n})$ such that $\boldsymbol{\mu_j}=(\mu_1^{(j)}, ... ,\mu_n^{(j)})$ where all $\mu_i^{(j)} \in \{0,1\}$, and the vector satisfies $\sum_{i} \mu_i^{(j)} = n-l$ if $j = v$ and $\sum_{i} \mu_i^{(j)} = l$ if $j \neq v$.
    
  \end{enumerate}
  
  \item \textbf{Mechanism phase.}
  \begin{enumerate}
    \item
    $\mathcal{V}$ picks random $a_j, b_j \leftarrow \mathbb{Z}_q$ and sends $g^{a_j}$, $g^{b_j}$ and $g^{a_jb_j-\sigma_j+1}$ to $\mathcal{P}$ for all $j\in [d]$.
    
    \item
    For all $j\in [d]$ and $i\in \{1,...,n\}$, $\mathcal{P}$ performs the following subroutine;
    (1) Generate $(r_i^{(j)}, s_i^{(j)})$ at random;
    (2) Compute $w_i^{(j)} \leftarrow g^{r_i^{(j)}}(g^{a_j})^{s_i^{(j)}} = g^{r_i^{(j)} + as_i^{(j)}}$ and $h_i^{(j)} \leftarrow (g^{b_j})^{r_i^{(j)}}(g^{a_jb_j-\sigma_j+1}g^{i-1})^{s_i^{(j)}} = g^{(r_i^{(j)}+as_i^{(j)})b_j+(i-\sigma_j)s_i^{(j)}}$;
    (3) Encrypt $\mu_i^{(j)}$ to $y_i^{(j)}$ as $y_i^{(j)} \leftarrow g^{\mu_i^{(j)}}h^{h_i^{(j)}}$.
    Then, $\mathcal{P}$ sends all ($w_i^{(j)}$, $y_i^{(j)}$) to $\mathcal{V}$.
    
    \item
    For all $j\in [d]$, $\mathcal{V}$ computes $g^{\mu_{\sigma_j}^{(j)}} \leftarrow y_{\sigma_j}^{(j)} / h^{(w_{\sigma_j}^{(j)})^{b_j}}$.
    And then, find $\mu_{\sigma_j}$.
    Thus, $\mathcal{V}$ receives $[\mu_{\sigma_1},...,\mu_{\sigma_d}]$ as a randomized output from $\mathcal{P}$.
    
  \end{enumerate}
  \item \textbf{Proof phase for P1.}
  \begin{enumerate}
    \item
    For all $j \in [d]$, for all $i \in \{1,...,n\}$, $\mathcal{P}$ generates challenge $c_{1-\mu_i^{(j)},i}^{(j)}$ and response $s_{1-\mu_i^{(j)},i}^{(j)}$ from $\mathbb{Z}_q$ and prepares commitments $com_{1-\mu_i^{(j)},i}^{(j)} \leftarrow h^{s_{1-\mu_i^{(j)},i}^{(j)}} / (y_i^{(j)} / g^{\mu_i^{(j)}})^{c_i^{(j)}}$.
    Generate $w_i^{(j)} \leftarrow \mathbb{Z}_q$ and compute $com_{(\mu_i^{(j)}),i}^{(j)} \leftarrow h^{w_i^{(j)}}$.
    Then, send $com_{\{0,1\}, i}^{(j)}$ to $\mathcal{V}$, for all $i, j$.
    
    \item
    $\mathcal{V}$ picks $x_i^{(j)} \leftarrow \mathbb{Z}_q$ for all $j\in [d]$ and $i\in {1,...,n}$ and sends it to $\mathcal{P}$.
    
    \item
    For all $j\in [d]$ and $i\in \{1,...,n\}$, $\mathcal{P}$ computes $c_{\mu_i^{(j)},i}^{(j)} = x_i^{(j)} - c_{1-\mu_i^{(j)},i}^{(j)}$ and $s_{\mu_i^{(j)},i}^{(j)} = v_i^{(j)}c_{\mu_i^{(j)},i}^{(j)} + w_i^{(j)}$.
    Then, send $c_{\{0,1\}, i}^{(j)}$ and $s_{\{0,1\}, i}^{(j)}$ for all $i, j$ to $\mathcal{V}$.
    
    \item
    Finally, $\mathcal{V}$ checks if $h^{s_{\{0,1\}, i}^{(j)}} = b(y_i^{(j)}/g^{\{0,1\}})^{c_{\{0,1\},i}^{(j)}}$ and $x_i^{(j)} = c_{0,i}^{(j)} + c_{1,i}^{(j)}$, for all $i\in \{1,...,n\}$ and for all $j\in [d]$.
    Otherwise halts.
  \end{enumerate}
  
  \item \textbf{Proof phase for P2. (Simplified because it is similar to P1.)}
  \begin{enumerate}
    \item 
    $\mathcal{P}$ generates and sends all $com_{\{p, q\}}^{(j)}$ to $\mathcal{V}$.
    
    \item
    $\mathcal{V}$ picks $x_j \leftarrow \mathbb{Z}_q$ for all $j\in [d]$ and sends it to $\mathcal{P}$.
    
    \item
    $\mathcal{P}$ sends $c_{\{p, q\}}^{(j)}$ and $s_{\{p, q\}}^{(j)}$ for all $j$ to $\mathcal{V}$
    
    \item
    $\mathcal{V}$ checks if $h^{s_{p}^{(j)}} = b(\prod_{i\in {1,..,n}} y_i^{(j)} / g^{n/2})^{c_{p}^{(j)}}$ and $h^{s_{q}^{(j)}} = b(\prod_{i\in {1,..,n}} y_i^{(j)} / g^{l})^{c_{q}^{(j)}}$ and $x_j = c_{p}^{(j)}+ c_{q}^{(j)}$ for all $j\in [d]$. Otherwise halts.
  \end{enumerate}
  
\item \textbf{Proof phase for P3.}
  \begin{enumerate}
    \item
    $\mathcal{P}$ computes $h_{sum} \leftarrow \sum_{i,j} h_i^{(j)}$ and sends $h_{sum}$ to $\mathcal{V}$.
    
    \item
    $\mathcal{V}$ checks if $h^{h_{sum}}g^{n/2+l(d-1)} = \prod_{i,j} y_i^{(j)}$. Otherwise halts.
  \end{enumerate}
\end{enumerate}
\end{protocol}

Then, similar to secure kRR, we must show that the all Bernoulli distributions represented by $d$ vectors are correct.
Specifically, the proofs are that all elements of bit vectors $\vec{\boldsymbol{\mu}}$ are surely a bit (0 or 1) (\textbf{P1}) and distribution of the vectors are surely equivalent to either of $p$ or $q$ of Eq. (\ref{eq:oue}) (\textbf{P2}) and the number of $p$ and $q$ are $1$ and $d-1$ respectively (\textbf{P3}).
If all these three proofs are verified, we can confirm the OUE protocol is simulated correctly.
Like kRR's proofs, \textbf{P1} and \textbf{P2} are proved by $d$ disjunctive proofs.
\textbf{P3} is based on hardness of discrete logarithm problem.
$\mathcal{P}$ cannot find $h_{sum}$ in polynomial time without all correct $h_i^{(j)}$ that is used when encrypting $y_i^{(j)}$.
While $\mathcal{P}$ has to release $h_{sum}$, this is information theoretically indistinguishable from $\mathcal{V}$ for each $h_i^{(j)}$ unless $n=d=1$.
Security statements for the secure OUE protocol are similar to secure kRR.
For LDP, as we can see 1.(a) in Protocol \ref{pro:oue}, we set $q=l/n$ such that $\frac{l}{n} \ge \frac{1}{1+e^{\epsilon}}$.

\subsection{Secure OLH}
To make OLH output-manipulation secure, basically we can use Protocol \ref{pro:krr} except that it requires sharing of a hash function and using reduced output category space.
As a first step, $\mathcal{V}$ generates and sends a seed $s$ to $\mathcal{P}$ to initialize hash function $H_{s}: v \rightarrow v'$ where $v \in [d]$ and $v' \in [g]$.
$\mathcal{V}$ and $\mathcal{P}$ use the same $H_{s}$ as a hash function.
We can apply Protocol \ref{pro:krr} to achieve secure OLH by using category set $[g]$ instead of $[d]$ and sensitive input value $v$ is handled as $v' = H_{s}(v)$.
The rest of the steps are almost the same as kRR.

Even if $\mathcal{P}^*$, who does not use the hash function correctly, participates the protocol, $\mathcal{V}$ can easily detect it if it sends the output of a different output space, i.e. $y\notin [g]$.
If attacker does not use a different output space, the attack can only be equivalent to input-manipulation because $\mathcal{V}$ verifies the correctness of the categorical distribution used in random sampling after applying hash function.

\section{Evaluation}
\label{sec:5}
In this section, we analyze our proposed protocol in terms of performance.

\noindent
\textbf{Experimental setup.} 
We use an HP Z2 SFF G4 Workstation, with 4-core 3.80 GHz Intel Xeon E-2174G CPU (8 threads, with 8MB cache), 64GB RAM.
The host OS is Ubuntu 18.04 LTS.
Our experimental implementation in Python is available on github\footnote{https://github.com/FumiyukiKato/verifiable-ldp}.
The client and server exchange byte data serialized by \textsf{pickle} (from Python standard library) over TCP.
We use $\epsilon=1.0$ and in OLH, set $g=d/2$ as the hashed space instead of $g=\lfloor e^{\epsilon} + 1 \rfloor$ for demonstration.

\begin{figure}[t]
\begin{minipage}{0.32\hsize}
    \centering
    \includegraphics[width=0.95\hsize]{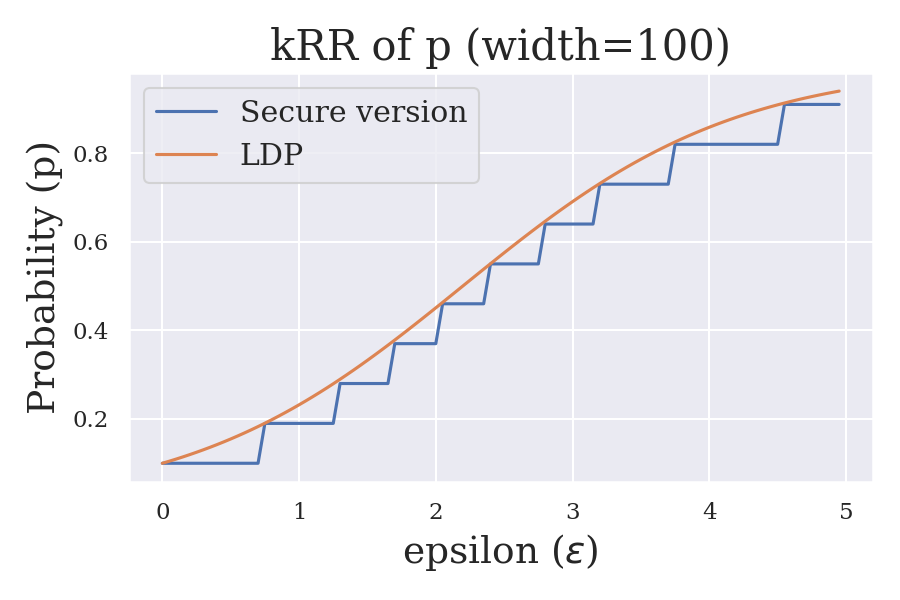}
\end{minipage}
\hfill
\begin{minipage}{0.32\hsize}
    \centering
    \includegraphics[width=0.95\hsize]{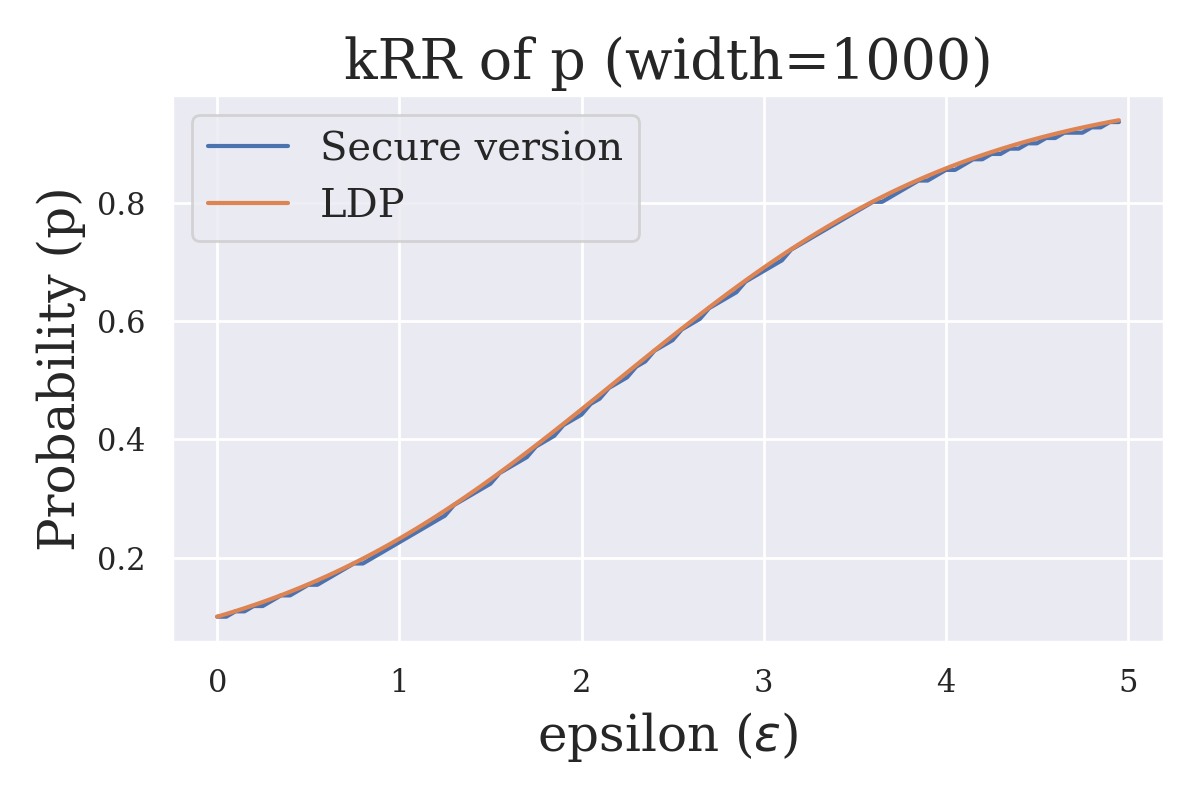}
\end{minipage}
\hfill
\begin{minipage}{0.32\hsize}
    \centering
    \includegraphics[width=0.95\hsize]{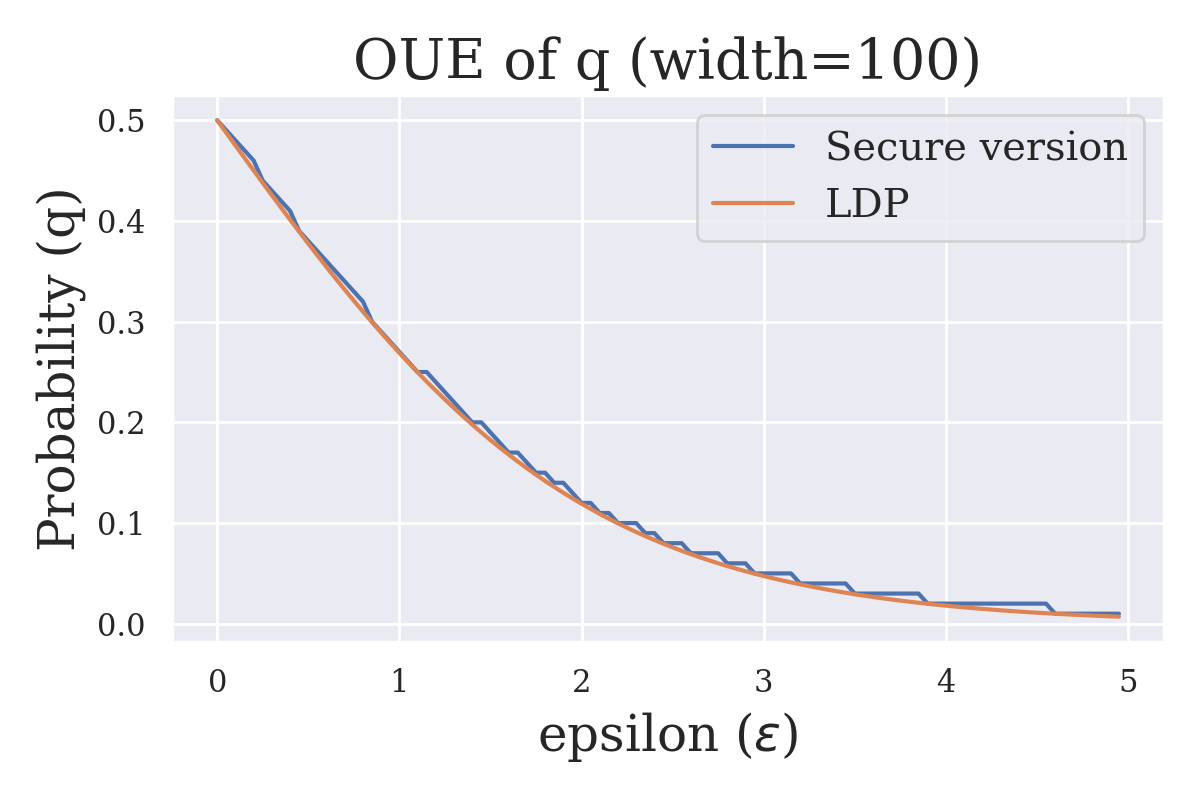}
\end{minipage}
\caption{In secure kRR, with a sufficiently large $width$, categorical distribution by Algorithm \ref{alg:params_krr} can accurately approximate the LDP distributions (left and middle). In secure OUE, it is almost exact discrete approximation with relatively small $width$. (right)}
\label{fig:dsp}
\vspace{-5mm}
\end{figure}

\noindent
\textbf{Parameter generator.}
First, we analyze the approximated probability distribution generated by our proposed method.
In secure kRR protocol, we approximate the probability distribution where we generate data to satisfy LDP by Algorithm \ref{alg:params_krr}.
Figure \ref{fig:dsp} shows how accurate the algorithm generates discrete distribution for $\epsilon=(0,5]$ and for $width=\{100, 1000\}$.
The red curve represents probability $p$ for the normal mechanism, and the blue one represents the approximated one.
When the $width$ is small, there is a noticeable loss of accuracy due to approximation.
However, with a sufficiently large $width$, the approximated $p$ has a sufficiently small loss.
As the $width$ increases, the performance degrades, indicating that there is a trade-off between the accuracy of the probability approximation and the performance.
This is true not only for kRR but also for OUE and OLH.
For secure OUE, in the right-side of Figure \ref{fig:dsp}, we compare probability $q$ because $p$ is constant in OUE.
It is almost exact discrete approximation with small $width$.
This is due to the difference in the structure of the vectors that form the probability distribution, with OUE having a simpler structure.

\begin{figure}[t]
\begin{minipage}{0.45\hsize}
    \centering
    \includegraphics[width=0.95\hsize]{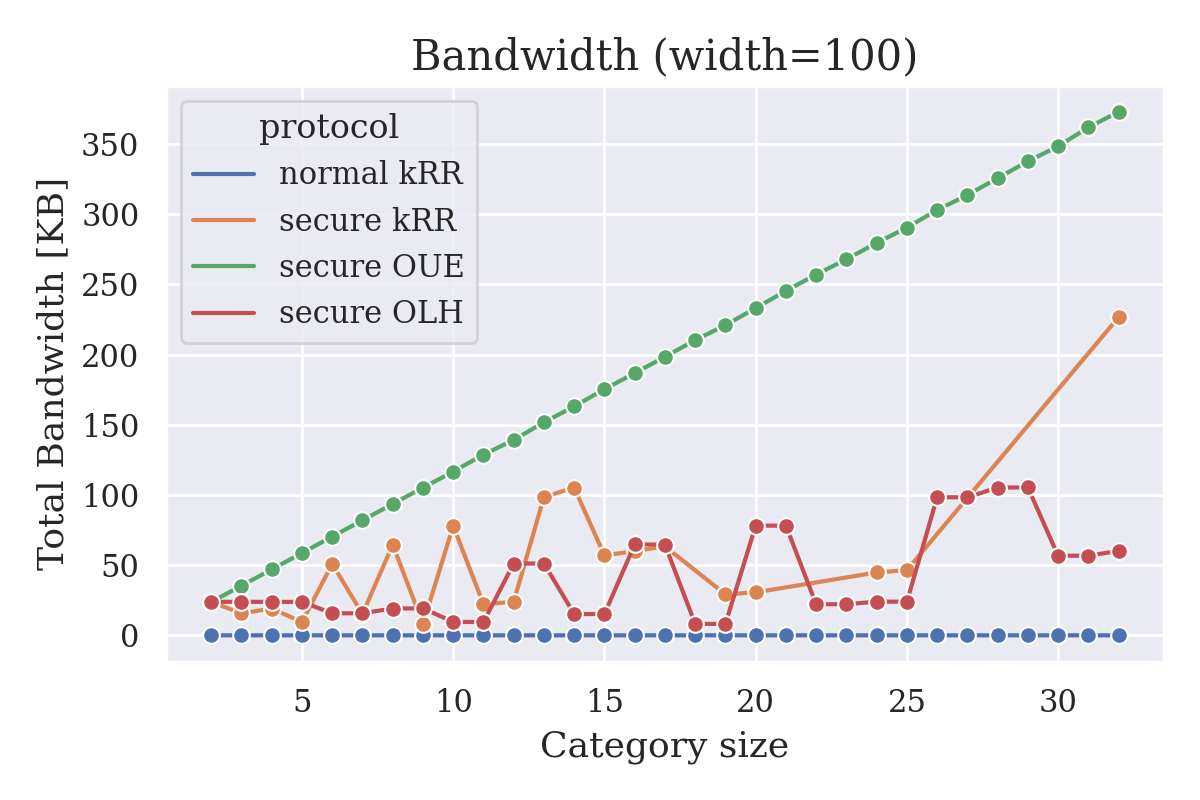}
\end{minipage}
\hfill
\begin{minipage}{0.45\hsize}
    \centering
    \includegraphics[width=0.95\hsize]{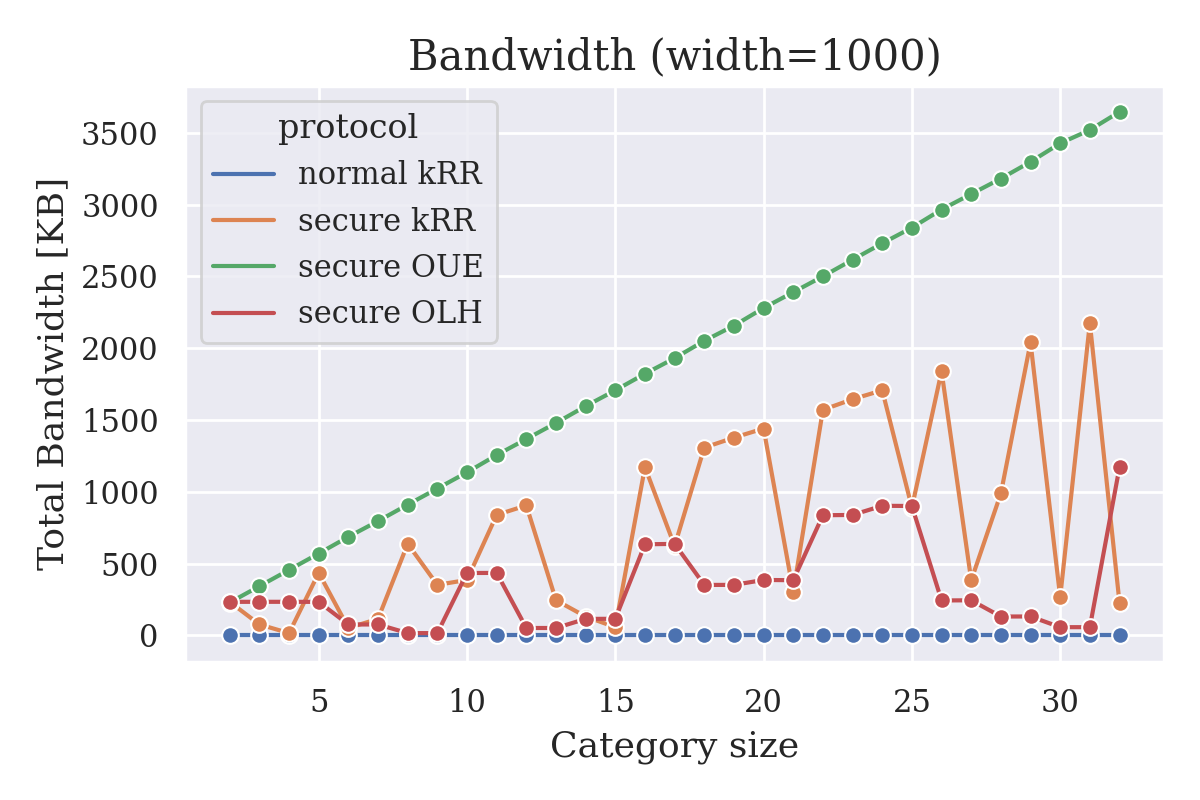}
\end{minipage}
\caption{With the same $width$, the communication costs of kRR and OLH are small. However, OUE can approximate LDP accurately with small widths (Fig. \ref{fig:dsp}).}
\label{fig:bandwidth}
\vspace{-3mm}
\end{figure}

\noindent
\textbf{Performance.}
We evaluate performances of our proposed method.
Figure \ref{fig:bandwidth} shows total bandwidths, caused in communications of the total protocol, of each three methods for different category sizes.
Generally, when increasing category size, total bandwidth also increases.
While it increases linearly in OUE, there are fluctuations in kRR and OLH.
This is because the probability value that Algorithm \ref{alg:params_krr} approximates may have a smaller denominator (i.e., $n$) by reduction, which can make the distribution vector smaller.
Overall, larger $width$ generates almost linear increases in bandwidth.
And for the same $width$, secure OUE causes larger communication overhead than others.
However, as mentioned in the previous paragraph, secure OUE can approximate the probability distribution with high accuracy using smaller $width$.
Hence, in particular, when the number of categories is large, secure OUE is considered to be more efficient by using smaller $width$.
Figure \ref{fig:bandwidth} shows that, comparing kRR with $width=1000$ and OUE with $width=100$, many categories require several times more bandwidth.
On the other hand, when the discretized probability distribution can be approximated with a small denominator by reduction, kRR and OLH show a very small bandwidth.
When comparing kRR and OLH, OLH is smaller overall.
This is due to the fact that the output space is reduced by hashing.

\begin{figure}[t]
\begin{minipage}{0.45\hsize}
    \centering
    \includegraphics[width=0.95\hsize]{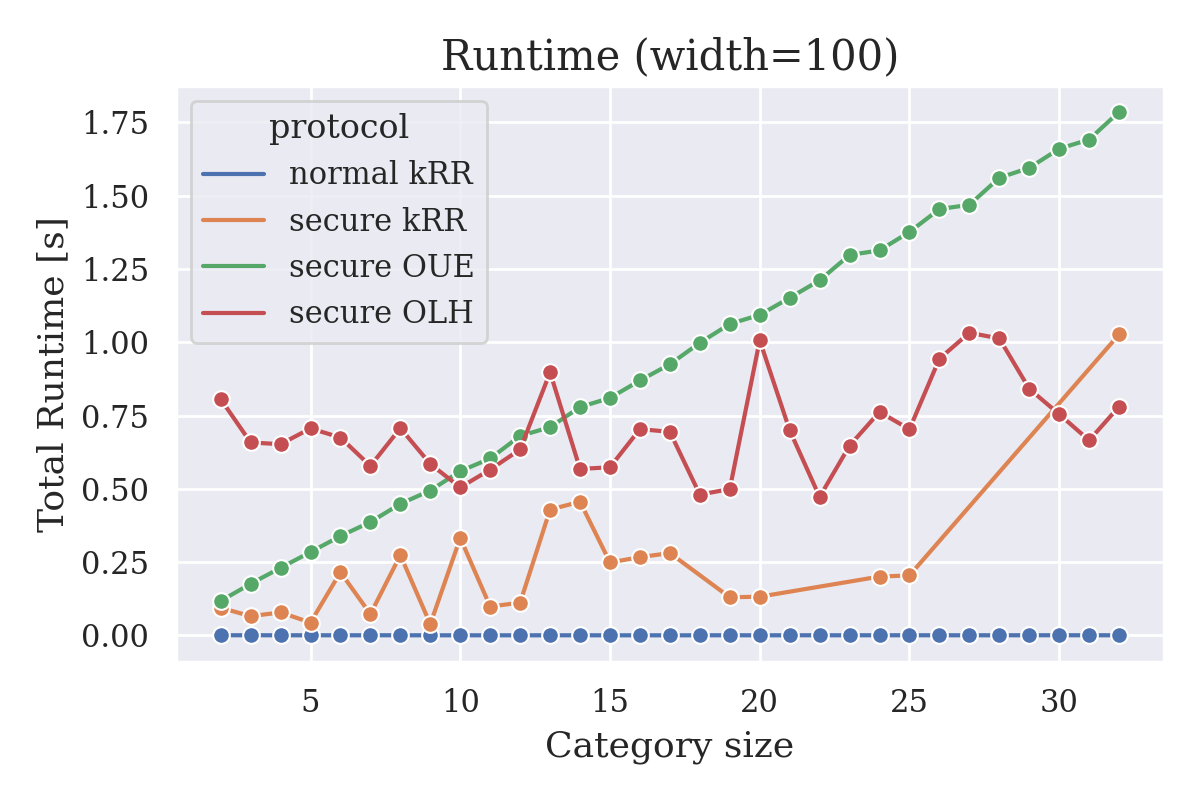}
\end{minipage}
\hfill
\begin{minipage}{0.45\hsize}
    \centering
    \includegraphics[width=0.95\hsize]{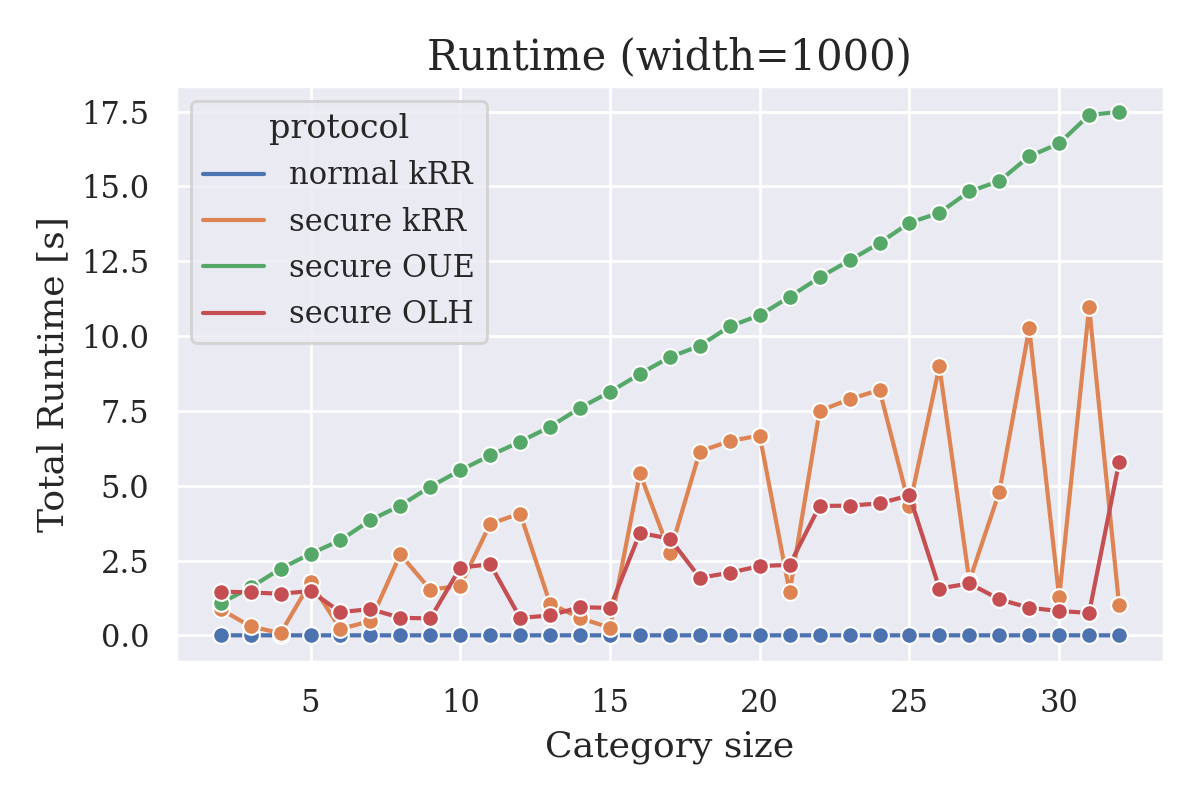}
\end{minipage}
\caption{The characteristics of runtime is similar to bandwidth. OLH takes a little longer because of the hashing.}
\label{fig:runtime}
\vspace{-3mm}
\end{figure}

Figure \ref{fig:runtime} shows total execution time from the time the client sends the first request until the entire protocol is completed.
Most of the characteristics are similar to those of bandwidth.
As the size of the proofs that need to be computed increases, the execution time is also expected to increase.
The only difference is OLH, which takes extra time to execute the hash function.
However, as the number of categories becomes larger, the influence becomes smaller.

Therefore, the overhead can be minimized by providing a privacy budget for optimal efficiency for kRR and OLH, and by using different methods for different $width$.
The overhead is expected to increase as the number of categories increases, but since the limit on the number of categories is determined to some extent by the use of LDP, we do not think this is a major problem.

At the end, impressively, our method is algorithm-only, making it more feasible than alternatives that assume secure hardware \cite{sgx} or TEE \cite{tee15}.
Nevertheless, overall, we believe the overhead is acceptable.
We believe this is due to the fact that we use relatively lightweight OT techniques as a building block.

\section{Conclusion}
\label{sec:6}
In this paper, we showed how we prevent malicious clients from attacking to LDP protocol.
An important observation was the effectiveness of output-manipulation and the importance of protection against it.
Our approach was verifiable randomization mechanism satisfying LDP.
Data collector can verify the completeness of executing agreed randomization mechanism for every possibly malicious data providers.
Our proposed method was based on only lightweight cryptography, hence, we believe it has high feasibility and can be implemented in various and practical data collection scenarios.

%
%
%
%

\end{document}